# Differences in chemical doping matter - Superconductivity in $Ti_{1-x}Ta_xSe_2$ but not in $Ti_{1-x}Nb_xSe_2$

Huixia Luo[1,*], Weiwei Xie[1], Jing Tao[2], Ivo Pletikosic[2,3], Tonica Valla[2], Girija S. Sahasrabudhe[1], Gavin Osterhoudt[4], Erin Sutton[4], Kenneth S. Burch[4] Elizabeth M. Seibel[1], Jason W. Krizan[1], Yimei Zhu[2] and Robert J. Cava[1,]

[1]*Department of Chemistry, Princeton University, Princeton, NJ 08544, USA*

[2]*Condensed Matter Physics and Materials Science Department, Brookhaven National Lab, Upton, New York 11973, USA*

[3]*Department of Physics, Princeton University, Princeton, NJ 08544, USA*

[4]*Department of Physics, Boston College, 140 Commonwealth Ave Chestnut Hill, Boston, MA 02467-3804, USA*

**ABSTRACT**

We report that 1T-TiSe$_2$, an archetypical layered transition metal dichalcogenide, becomes superconducting when Ta is substituted for Ti but not when Nb is substituted for Ti. This is unexpected because Nb and Ta should be chemically equivalent electron donors. Superconductivity emerges near $x = 0.02$ for Ti$_{1-x}$Ta$_x$Se$_2$, while for Ti$_{1-x}$Nb$_x$Se$_2$, no superconducting transitions are observed above 0.4 K. The equivalent chemical nature of the dopants is confirmed by X-ray photoelectron spectroscopy. ARPES and Raman scattering studies show similarities and differences between the two systems, but the fundamental reasons why the Nb and Ta dopants yield such different behavior are unknown. We present a comparison of the electronic phase diagrams of many electron-doped 1T-TiSe$_2$ systems, showing that they behave quite differently, which may have broad implications in the search for new superconductors. We propose that superconducting Ti$_{0.8}$Ta$_{0.2}$Se$_2$ will be suitable for devices and other studies based on exfoliated crystal flakes.

*huixial@princeton.edu

## INTRODUCTION

Layered transition-metal dichalcogenides (TMDs) have been studied for decades as archetypical examples of materials where superconductivity is balanced against a competing charge density wave (CDW) state.[1-10] The superconducting and CDW transition temperatures in this family can be tuned by changing electron count through chemical substitution or intercalation (e.g. refs. 11-13), using high pressure (e.g. refs. 14-17), or gating[18]). 1T-TiSe$_2$ is one of the simplest and most widely studied TMDs, undergoing a transition to a CDW state at about 200 K in its native form[19] and becoming a superconductor when put under pressure or electron doped through intercalation. In Cu$_x$TiSe$_2$, Cu donates electrons to the TiSe$_2$ layers, and superconductivity is induced with a maximum T$_c$ of 4.2 K. This observation has triggered a great deal of recent activity on long-studied 1T-TiSe$_2$ (e.g. refs 13, and 19-24), especially as the superconducting phase is proposed to be an example of an exciton condensate. Similarly, Pd-intercalated TiSe$_2$ is also superconducting.[24]

Here we report the observation of superconductivity in 1T-TiSe$_2$ induced by doping with electrons through partial substitution of Ta for Ti, in materials of the type Ti$_{1-x}$Ta$_x$Se$_2$. We find that for Ti$_{1-x}$Ta$_x$Se$_2$ the CDW transition remains present and a superconducting state emerges near $x = 0.02$ with a maximum T$_c$ of 2.2 K at $x = 0.2$. In contrast, we find that similarly made and tested isostructural and chemically isoelectronic Ti$_{1-x}$Nb$_x$Se$_2$ is not superconducting above 0.4 K. This is unexpected because both Nb and Ta have 5 valence electrons, and thus should simply donate their electrons to the conduction band of 1T-TiSe$_2$, which is dominated by normally empty Ti (4 valence electrons) electronic states. This conventional electronic picture is verified by our chemical spectroscopy (X-ray photoelectron spectroscopy (XPS)) measurements, and our ARPES characterization of the materials shows that electrons are indeed donated to the formerly empty conduction band in 1T-TiSe$_2$ by both substitutions, but also that there are some significant differences. Consistent with the ARPES characterization, the Nb substitution leads to a lower electronic density of states than the Ta substitution, inferred from specific heat measurements. Further the Nb substituted material shows non-metallic resistivity behavior, in contrast to the metallic and superconducting behavior induced by Ta substitution. Finally, we construct a composition-dependent superconductivity phase diagram for many dopants in the archetype 1T-TiSe$_2$ system, comparing Ti$_{1-x}$Ta$_x$Se$_2$, Ti$_{1-x}$Nb$_x$Se$_2$, Pd$_x$TiSe$_2$ and Cu$_x$TiSe$_2$. The phase diagram shows that the superconductivity that is induced in doped 1T-TiSe$_2$ is dramatically dependent on the chemical method used to change its electron count. This result for the TMD 1T-TiSe$_2$ is in contrast to what is found for other important superconducting systems, such as the iron arsenides, where substitutions of many different kinds induce nearly equivalent maximum superconducting T$_c$'s at the same electron count.[25-27] Our results show that what appear to be chemically equivalent electron donors are in fact not at all electronically equivalent in this system. If this is frequently the case, then it raises significant general issues in the search for superconductivity in all doped materials, where chemically equivalent dopants are only rarely individually tested.

**EXPERIMENTAL SECTION**

Polycrystalline samples of Ti$_{1-x}$Ta$_x$Se$_2$ and Ti$_{1-x}$Nb$_x$Se$_2$ were synthesized in two steps by solid state reaction. First, the mixtures of high-purity fine powders of Ta (99.8%) or Nb (99.8%), Ti (99.9%) and Se (99.999%) in the appropriate stoichiometric ratios were thoroughly ground, pelletized and heated in sealed evacuated silica tubes at a rate of 1 °C/min to 700 °C and held there for 120 h. Subsequently, the as-prepared powders were reground, re-pelletized, and sintered again, heated at a rate of 3 °C/min to 700 °C and held there for 120 h. Single crystals of selected compositions were grown by the chemical vapor transport (CVT) method with iodine as a transport agent. Two- gram as-prepared powders of Ti$_{0.8}$Ta$_{0.2}$Se$_2$ or Ti$_{0.8}$Nb$_{0.2}$Se$_2$ were mixed with 100 mg iodine, sealed in evacuated silica tubes and heated for one week in a two zone furnace, where the temperature of source and growth zones were fixed at 675 °C and 725 °C, respectively. The identity and phase purity of the samples were determined by powder X-ray diffraction (PXRD) using a Bruker D8 Advance ECO with Cu Kα radiation and a LYNXEYE-XE detector. To determine the unit cell parameters, LeBail fits were performed on the powder diffraction data through the use of the FULLPROF diffraction suite using Thompson-Cox-Hastings pseudo-Voigt peak shapes.[28] Single crystals selected from partially crushed crystalline samples were employed for the single crystal structure determinations.

Measurements of the temperature dependence of the electrical resistivity (4 contact), specific heat and magnetic susceptibility of the materials were performed in a Quantum Design Physical Property Measurement System (PPMS). There was no indication of air-sensitivity of the materials during sample preparation. Selected resistivities and heat capacities were measured in the PPMS equipped with a $^3$He cryostat. Magnetic susceptibility characterization for $Ti_{0.8}Ta_{0.2}Se_2$ and $Ti_{0.8}Nb_{0.2}Se_2$ was carried out in a 5T applied AC field. Specimens for the electron diffraction studies in a transmission electron microscope were obtained from synthesized samples crushed in a dry box and transported to the microscope in ultra-high vacuum. Temperature-dependent electron diffraction measurements were performed at Brookhaven National Laboratory on a JEOL 2100F microscope equipped with a liquid-helium cooled sample holder. The angle-resolved photoelectron spectroscopy (ARPES) measurements were conducted at beamlines 10 and 12 of Advanced Light Source, Lawrence Berkeley National Laboratory using Scienta electron analyzers set to overall resolution of 25 meV and 0.3°. Two-dimensional angular maps were assembled at BL10 from multiple line scans taken by rotating the analyzer around the axis parallel to its slit. Samples were cleaved at 15 K in ultrahigh vacuum of $5 \times 10^{-9}$ Pa and all the data were collected at 15 K. The phonon spectra of Nb- and Ta- doped $TiSe_2$ were probed using micro-Raman spectroscopy. In layered TMDs this can be challenging due to their strong tendency to oxidize at the surface. Thus we performed the experiments entirely in a glovebox with argon atmosphere, with samples being freshly cleaved just before the measurement. This was achieved with a WITec alpha300R spectrometer customized to work inside an Ar-filled glovebox. The sample was excited with unpolarized light at 532 nm with the reflected and Raman scattered light collected in a backscattering configuration. The reflected light was removed using an edge filter, resulting in a lower cut-off of 85 cm$^{-1}$. To avoid unwanted heating, the power was kept below 20 μW and focused to a spot size approximately 1 μm in diameter. Results shown are the average of at least 6 such measurements, corrected for the integration time and laser power. To confirm the single crystal nature and reproducibility, all spectra were confirmed by measuring spots millimeters apart.

X-ray Photoelectron Spectroscopy (XPS) characterization was performed with a VG ESCA Lab Mk.II instrument. All spectra were obtained using Mg Kα radiation (1284 eV) and 20 eV pass energy. $NbSe_2$, $Ti_{0.8}Nb_{0.2}Se_2$, $Ti_{0.8}Ta_{0.2}Se_2$ polycrystals and $TaSe_2$ single crystals were placed on carbon tape attached to separate metal sample holders. Usually, the Carbon 1s (C1s) peak originating from adventitious carbon on the sample surface is used for calibration purposes. But as the samples were polycrystalline, C1s signal from the carbon tape could not be obviated. Thus, to compensate for the charging effects, the sample holders were biased at +10 volts [29].

Since, the surface of the polycrystals and single crystals were oxidized due to ambient oxidation; $TiO_2$, $Nb_2O_5$, $Ta_2O_5$ formed at the surface of the samples were used for comparison and calibration. All scans were taken with a 0.05 eV step size and 0.5 s dwell time. The resolution of the instrument is less than 0.1 eV. The obtained scans were fit with Casa XPS using a Shirley background, area and positions were constrained using standard values.

**Results and Discussion**

First we consider the chemistry and structures of the $Ti_{1-x}Ta_xSe_2$ and $Ti_{1-x}Nb_xSe_2$ systems. 1T-$TiSe_2$ is a layered compound with trigonal symmetry.[30] The Ti atoms, which are in octahedral coordination with Se, form planar $TiSe_2$ layers of edge sharing octahedra. These layers are bonded to each other by van der Waals forces. Previous work has shown that when Cu atoms are intercalated to form the $Cu_xTiSe_2$ superconductor, they occupy positions between the $TiSe_2$ layers.[13] Here we find from our high quality single crystal structural analyses of $Ti_{0.9}Ta_{0.1}Se_2$ and $Ti_{0.8}Nb_{0.2}Se_2$ that when Ta or Nb atoms are substituted for Ti, they substitute directly on the Ti site, replacing some of the Ti in the octahedra. There are no interstitial atoms in either case, to a high level of precision, and both structures are that of ideal 1T-$TiSe_2$ (see **Table 1S** and **Table 2S** Supplementary Information). **Figure 1a** shows the powder x-ray diffraction patterns for selected members of both families. The results show that single phase solid solutions are indeed formed in these systems. The solubility limit for intercalated Cu in $TiSe_2$ is $x \approx 0.11$. However, in the substitution case, the solubility limits for $Ti_{1-x}Ta_xSe_2$ and $Ti_{1-x}Nb_xSe_2$ in the 1T structure phase are $x \approx 0.9$ and $x \approx 0.7$, respectively; at higher doping contents, the 2H-type TMDC structure is found for both $Ti_{1-x}Ta_xSe_2$ and $Ti_{1-x}Nb_xSe_2$.

The composition dependence of the room temperature lattice parameters for 1T-$Ti_{1-x}Ta_xSe_2$ ($0 \leq x \leq 0.9$), 1T-$Ti_{1-x}Nb_xSe_2$ ($0 \leq x \leq 0.7$), and a comparison to those for 1T-$Cu_xTiSe_2$ ($0 \leq x \leq 0.11$) are shown in **Figure 1b**. The *a* parameters increase through both substitution of Ta or Nb and intercalation of Cu in $TiSe_2$, but the *c* parameters change in an opposite fashion for substitution vs. intercalation: *c* decreases with increasing Ta or Nb substitution in $Ti_{1-x}(Ta/Nb)_xSe_2$, while it increases with increasing Cu intercalation in $Cu_xTiSe_2$ ($0 \leq x \leq 0.11$). The fact that the lattice parameters track each other so well in the two cases is an indirect indication that the

Nb and Ta doped systems are structurally analogous. The anomalous *c* axis behavior of $Cu_xTiSe_2$ has been previously noted.[13]

Superconductivity emerges near $x = 0.02$ for $Ti_{1-x}Ta_xSe_2$, while for $Ti_{1-x}Nb_xSe_2$, no superconducting transitions are observed above 0.4 K in the broad composition range of $0 \leq x \leq 0.7$. Looking to find differences in the chemistry of two systems, we performed XPS studies $Ti_{0.8}Ta_{0.2}Se_2$ and $Ti_{0.8}Nb_{0.2}Se_2$, as shown in **Figure 1 e, f**. For comparison, the Nb *3d* and Ta *4f* spectra for undoped 2H-NbSe$_2$ and 2H-TaSe$_2$ are included in **Figure 1 c, d**. The binding energy of the Ta $4f_{7/2}$ peak in TaSe$_2$ is 0.8 eV lower than that in $Ti_{0.8}Ta_{0.2}Se_2$. The binding energy of the Ta $4f_{7/2}$ peaks corresponding to $Ta_2O_5$ formed at the surface of TaSe$_2$ and $Ti_{0.8}Ta_{0.2}Se_2$ is 26.5 eV [31] Similarly, the binding energy of the Nb $3d_{5/2}$ peak in NbSe$_2$ is 1.2 eV lower than that in $Ti_{0.8}Nb_{0.2}Se_2$. The binding energy of the Nb $3d_{5/2}$ peaks corresponding to $Nb_2O_5$ formed at the surface of NbSe$_2$ and $Ti_{0.8}Nb_{0.2}Se_2$ is 207.5 eV [32] Thus, both Ta and Nb are more oxidized (i.e. have formal oxidation states between 4+ and 5+) when used as dopants in TiSe$_2$ than in the individual selenides. The relative shifts in binding energies are the same for both species, indicating that as chemical dopants they are indeed equivalent in 1T-TiSe$_2$. The Ti *2p* and Se *3d* XP spectra for both $Ti_{0.8}Ta_{0.2}Se_2$ and $Ti_{0.8}Nb_{0.2}Se_2$ are identical [33] as shown in **Figures 1S and 2S** (Supporting Information) further supporting the chemical equivalence of the two systems

We next consider the transport properties of the two systems. A systematic change in the temperature dependence of the resistivity of $Ti_{1-x}Ta_xSe_2$ occurs on increasing *x*. **Figures 2a** shows the temperature dependence of the normalized electrical resistivity ($\rho/\rho_{300K}$) for polycrystalline samples of $Ti_{1-x}Ta_xSe_2$ ($0 \leq x \leq 0.3$). At low temperatures, a clear, sharp ($\Delta T_c < 0.1$ K) drop of $\rho(T)$ is observed in the doped samples, signifying the onset of superconductivity at low temperatures in $Ti_{1-x}Ta_xSe_2$ for $x > 0.02$; as the $Ti_{1-x}Ta_xSe_2$ compounds become better metals, superconductivity emerges. The Ta substituted sample with $x = 0.2$ shows the highest $T_c$, 2.2 K (inset of **Figure 2a**). In addition, the signature of the CDW transition is seen for the low *x* content samples through the presence of the maxima in $\rho(T)$; at higher doping content the signature of the CDW transition gets much weaker.

The temperature dependence of the normalized electrical resistivities ($\rho/\rho_{300K}$) for the polycrystalline samples of $Ti_{1-x}Nb_xSe_2$ ($0 \leq x \leq 0.7$) are shown in **Figure 2b.** In contrast to the situation for $Ti_{1-x}Ta_xSe_2$, non-metallic behavior is clearly observed. We examine the non-metallic behavior more closely in **Figure 2c**. The figure shows that the low temperature data can be fit by a two-dimensional variable range hopping model $\rho(T) = \rho_0 \exp(T_0/T)^n$, where $T_0$ is the characteristic Mott temperature, which depends on the electronic structure, the density of states near the Fermi level and localization length, $\rho_0$ is the pre-exponential factor and $n = 1/(d+1)$ for d-dimensional variable range hopping.[34] The materials are clearly not semiconducting at low temperatures, for which n = 1, although at higher temperatures the behavior appears to be semiconducting, with an activation energy of $E_A$ = 0.17 eV. No superconducting transition is seen in any of the Nb substituted samples down to 0.4 K.

Hall measurement data confirms that both the $Ti_{1-x}Ta_xSe_2$ and $Ti_{1-x}Nb_xSe_2$ materials are *n*-type as expected for electron doping of 1T-TiSe$_2$; the larger negative Hall resistivity and its increase in magnitude with decreasing temperature for $Ti_{0.8}Nb_{0.2}Se_2$ (**Figure 2d inset**) is consistent with a lower n-type carrier concentration than in the Ta doping case. Further, **Figure 2d** shows that in neither case does the substitution in 1T-TiSe$_2$ lead to localized magnetic states; induced magnetism being a possible reason for the differences in behavior for the two systems. The susceptibilities are diamagnetic, dominated by the core diamagnetism, and the small Curie tails at low temperatures are from a very small fraction (sub percent) of spin-bearing defects. Thus magnetism induced by doping cannot be behind the difference in the electronic behavior observed in the two systems.

We next consider a comparison of the low temperature specific heats of the two systems and the thermodynamic characterization of the new superconductor. **Figure 2e** shows the specific heat data employed in order to further investigate the electronic properties and superconductivity in the optimal $Ti_{1-x}Ta_xSe_2$ superconductor. The main panel of **Figure 2e** shows the temperature dependence of the specific heat ($C_p/T$ versus $T^2$) under zero-field and under 5 Tesla field for $Ti_{0.8}Ta_{0.2}Se_2$. For comparison, the temperature dependence of the zero-field specific heat ($C_p/T$ versus $T^2$) for $Ti_{0.8}Nb_{0.2}Se_2$ is shown in **Figure 2f**. In both materials, the specific heat at low temperatures (but above $T_c$) obeys the relation of $C_p = \gamma T + \beta T^3$, where $\gamma$ and $\beta$ describe the electronic and phonon contributions to the heat capacity, respectively, the latter of which is a measure of the Debye Temperature ($\theta_D$), and the former of which is the Sommerfeld parameter. By fitting the data in the temperature range of 2 - 10 K, we obtain the electronic specific heat coefficient $\gamma$ = 1.99 mJ·mol$^{-1}$·K$^{-2}$, and the phonon specific heat coefficient $\beta$ = 0.701 mJ·mol$^{-1}$·K$^{-4}$ for Ti$_{0.8}$Ta$_{0.2}$Se$_2$. Fitting the data for Ti$_{0.8}$Nb$_{0.2}$Se$_2$ similarly yields $\gamma$ = 0.45 mJ·mol$^{-1}$·K$^{-2}$ and $\beta$ = 0.475 mJ·mol$^{-1}$·K$^{-4}$. We can estimate the Debye temperatures by using the values of $\beta$, and $\theta_D$ = $(12\pi^4 nR/5\beta)^{1/3}$, where n is the number of atoms per formula unit (n = 3), and R is the gas constant. The $\theta_D$ values are thus calculated to be 202 K for Ti$_{0.8}$Ta$_{0.2}$Se$_2$ and 230 K for Ti$_{0.8}$Nb$_{0.2}$Se$_2$. Finally, it can be seen that $\gamma$ in Ti$_{0.8}$Ta$_{0.2}$Se$_2$ is nearly 5 times of that of Ti$_{0.8}$Nb$_{0.2}$Se$_2$. Since the value of $\gamma$ is proportional to the electronic density of states (DOS) near the Fermi level ($E_F$), and the DOS near $E_F$ has a very strong influence on $T_c$, this difference is likely a major factor in the lack of a superconducting transition in the Nb case. These data do not, however, tell us why the nominally equivalent Nb doping and Ta doping of 1T-TiSe$_2$ yield such different $\gamma$s.

Ti$_{0.8}$Ta$_{0.2}$Se$_2$ displays a large specific heat jump associated with a transition to superconductivity at $T_c$, as shown in the insets for **Figures 2e and f**. The superconducting transition temperature observed in the specific heat measurements for Ti$_{0.8}$Ta$_{0.2}$Se$_2$ is in excellent agreement with the $T_c$ determined in the $\rho(T)$ measurements. From the inset in **Figure 2a**, using the equal area construction method, we obtain $\Delta C/T_c$ = 3.78 mJ mol$^{-1}$ K$^{-2}$ for Ti$_{0.8}$Ta$_{0.2}$Se$_2$. The normalized specific heat jump value $\Delta C/\gamma T_c$ is thus found to be 1.9 for Ti$_{0.8}$Ta$_{0.2}$Se$_2$, somewhat higher than that of the Bardeen-Cooper-Schrieffer (BCS) weak-coupling limit value (1.43), confirming bulk superconductivity. Using the Debye temperature ($\theta_D$), the critical temperature $T_c$, and assuming that the electron-phonon coupling constant ($\lambda_{ep}$) can be calculated from the inverted McMillan formula[35]: $\lambda_{ep} = \dfrac{1.04 + \mu^* \ln\left(\dfrac{\theta_D}{1.45 T_C}\right)}{(1-0.62\mu^*)\ln\left(\dfrac{\theta_D}{1.45 T_C}\right) - 1.04}$, the value of $\lambda_{ep}$ obtained is 0.61 for Ti$_{0.8}$Ta$_{0.2}$Se$_2$. This suggests weak coupling superconductivity. The density of states at the Fermi level ($N(E_F)$) can be calculated from $N(E_F) = \dfrac{3}{\pi^2 k_B^2 (1+\lambda_{ep})} \gamma$ by using the value of $\gamma$ and the electron-phonon coupling ($\lambda_{ep}$). This yields $N(E_F)$ = 0.53 states/eV f.u. for this system's optimal superconductor Ti$_{0.8}$Ta$_{0.2}$Se$_2$.

The superconducting transition for the optimal superconducting sample was further examined through temperature dependent measurements of the electrical resistivity under applied magnetic field. The $\rho(T,H)$ obtained for Ti$_{0.8}$Ta$_{0.2}$Se$_2$ is presented in the supplementary information, **Figure 3S**. Based on the $T_c$ determined for different magnetic fields, the upper critical field values, $\mu_0 H_{c2}$, are plotted vs. temperature in the inset to **Figures 3S**. A clear linear dependence of $\mu_0 H_{c2}$ vs. T is seen near $T_c$; the solid line through the data shows the best linear fit with the initial slope $dH_{c2}/dT$ = -1.4 T/K for both $Ta_{0.2}Ti_{0.8}Se_2$ and. $Ta_{0.15}Ti_{0.85}Se_2$. We estimate the zero temperature upper critical field $\mu_0 H_{c2}$ = 2.23 T for $Ti_{0.8}Ta_{0.2}Se_2$ (and 2.21 T for $Ta_{0.15}Ti_{0.85}Se_2$) using the Werthamer-Helfand-Hohenberg (WHH) expression, $\mu_0 H_{c2}$ = -0.693$T_c$ $(dH_{c2}/dT_c)^{36-38}$. The upper critical field $\mu_0 H_{c2}(0)$ calculated for $Ti_{0.8}Ta_{0.2}Se_2$ is larger than that reported for the $Cu_{0.08}TiSe_2$, ($T_c$ = 4.15 K, $\mu_0 H_{c2}(0)$ = 1.33 T).[13] From $\mu_0 H_{c2} = \dfrac{\phi_0}{2\pi\xi_{GL}^2}$, where $\phi_o$ is the quantum of flux, the Ginzburg-Laudau coherence length can be estimated as $\xi_{GL}(0) \sim 120$ Å for $Ti_{0.8}Ta_{0.2}Se_2$.

Returning to the comparison of the two systems, we consider their characterization by low temperature electron diffraction, which is an excellent probe of the existence of CDWs in layered dichalcogenides.[9] Thus in **Figures 1g-j** we compare the electron diffraction patterns in the [001] diffraction zones for both $Ti_{0.8}Ta_{0.2}Se_2$ and $Ti_{0.8}Nb_{0.2}Se_2$, determined in the TEM experiments at both ambient temperature and 89 K, the latter temperature chosen to be low enough to probe the possible presence of a CDW. Through these patterns we can determine whether the presence of superconductivity in The Ta one case but not in the Nb case has to do with whether the CDW is more efficiently suppressed through the doping, thus tipping the CDW-superconductivity balance toward the latter. The results are initially surprising. They show that the (½, ½, ½) superlattice due to the CDW is very weak or absent at 89 K in non-superconducting $Ti_{0.8}Nb_{0.2}Se_2$ but is clearly present in superconducting $Ti_{0.8}Ta_{0.2}Se_2$. Thus the appearance of superconductivity in the Ta-doped case cannot be due to a more efficient suppression of the CDW by Ta doping. The CDW in $TiSe_2$, however, is far from conventional in character and the literature remains divided on its origin. [13-15] Therefore in $TiSe_2$, at least, whether the existence of the CDW should exclude the presence of superconductivity should not a priori be expected, and in fact is clearly not the current case. The interesting electronic picture for electron-doped doped 1T-$TiSe_2$ is further elaborated through our ARPES characterization of the electronic structures of $Ti_{0.85}Nb_{0.15}Se_2$ and $Ti_{0.85}Ta_{0.15}Se_2$, described below.

In **Figures 3a-h** we present the electronic structures of $Ti_{0.85}Ta_{0.15}Se_2$, $Ti_{0.85}Nb_{0.15}Se_2$ and pristine 1T-TiSe$_2$ determined in the ARPES experiments, which were performed at 15 K. Fermi surface cuts at the border of the Brillouin zone in the plane containing the high-symmetry points A, L, and H, at $k_c \sim \pi/c$, are shown in **Figure 3a** for $Ti_{0.85}Ta_{0.15}Se_2$ and **3b** for $Ti_{0.85}Nb_{0.15}Se_2$. The cuts show the petal-like electron Fermi surfaces from the conduction bands, analogous to what is seen in $Cu_xTiSe_2$.[39] The direct comparison shows the qualitatively smaller electron Fermi surface for the case of Nb doping, even though the chemically equivalent dopants are expected to be electronically identical as well. Panels **3c** and **3d** show the band dispersions across the electron pocket at L for $Ti_{0.85}Ta_{0.15}Se_2$ and $Ti_{0.85}Nb_{0.15}Se_2$, demonstrating the similarity in the dispersions, but again illustrating the smaller filling of the electron pocket in the Nb-doping case. Estimates of the n-type carrier concentrations from the sizes of the Fermi surfaces are $\sim 1 \times 10^{21}$ cm$^{-3}$ for $Ti_{0.85}Nb_{0.15}Se_2$ and $4 \times 10^{21}$ cm$^{-3}$ for $Ti_{0.85}Ta_{0.15}Se_2$, ARPES was used to study the character of the top of the valence band, that is, the bands forming the hole pockets in the center of the $k_c$-projected Brillouin zone for **3e** $Ti_{0.85}Ta_{0.15}Se_2$ and **3f** $Ti_{0.85}Nb_{0.15}Se_2$. These bands exhibit a noticeable reduction of the spectral intensity approximately 100 mev below the Fermi level. Some calculate this to be the signature of a CDW phase with moderate to strong excitonic effects.[36]

A general comparison between the cases of 1T-TiSe$_2$ and Nb-doped TiSe$_2$ is shown in panels **3g** and **3h**. The results for 1T-TiSe$_2$ **3g** show what is so unusual about the electronic structure of this material – the band folding due to the CDW is reflected in the fact that at the M point in the Brillouin zone the valence band and the conduction band almost "touch" at $E_F$ with an electronic deformation (i.e. deviation from simple parabolic behavior) at the bottom of the conduction band.[33] Thus the low temperature electronic structure of 1T-TiSe$_2$ is not analogous to what is seen for the "Fermi surface nesting" scenario displayed by other layered TMDCs with CDW transitions, such as NbSe$_2$.[40] Comparison of the 1T-TiSe$_2$ electronic structure (**3g**) to the case of the Nb doping (**3h**) shows that, as expected, the electrons donated by Nb result in significant occupancy of the conduction band. Just like pristine TiSe$_2$, the doped samples show the hole-like band replicated below the electron pocket at M - however, with considerably lower spectral intensity. We note finally that there is a considerably larger energy overlap between the hole-like bands around Γ(A) and the electron pockets around M(L) in Ta- and Nb- doped TiSe$_2$ than in either pristine 1T-TiSe$_2$, or 1T-TiSe$_2$ intercalated with Cu.[42]

Because superconductivity ultimately arises from electron-phonon coupling in conventional materials, we look further into the potential differences between the doped systems by comparing their phonon spectra, probed by Raman scattering, to that of undoped TiSe$_2$. The Raman spectra for 1T-TiSe$_2$ and the 15% Nb and Ta doped samples are shown in **Figure 3i**. The 1T-TiSe$_2$ Raman spectrum is in good agreement with previously published studies[43,44]. Specifically, we observe a strong $A_{1g}$ peak at 200 cm$^{-1}$ and an $E_g$ peak at 136 cm$^{-1}$ (the symmetries were established in previous studies). The Nb-doped sample produces a near identical spectrum to that of undoped TiSe$_2$. Interestingly, the $E_g$ mode is unaffected by Ta-doping, while two significant differences are observed near the $A_{1g}$ mode. Specifically, the $A_{1g}$ mode shifts to lower energies, while a new mode appears above it. This is best seen in **Figure 3j** where we focus on just the region near the $A_{1g}$ mode. By fitting with two Lorentzians, we find that the $A_{1g}$ mode has been shifted down to ~197 cm$^{-1}$ while a new mode has appeared at ~213 cm$^{-1}$. The shift of the $A_{1g}$ mode to lower energies is consistent with previous studies of 1T-TaSe$_2$, where the mode is found at 190 cm$^{-1}$ with no others in this range [45]. 1T-TaSe$_2$ has only been measured in its commensurate CDW state. Nonetheless, from group theory, we would not expect an additional mode in the absence of a CDW distortion. 2H-TaSe$_2$ does possess a mode very close to the observed new mode, but 2H-TaSe$_2$ could not be present as a separate phase because it would display two additional modes in the studied frequency range (at 210 cm$^{-1}$ and at 240 cm$^{-1}$) [46]. Given the high doping levels in Ti$_{0.8}$Ta$_{0.2}$Se$_2$ this could instead be a local defect induced mode resulting from the Ta doping. Ta is quite a bit heavier than Ti, however and as such is expected to produce local modes below the bulk modes and not above as is observed here. Ultimately further studies using polarization and/or temperature dependence could potentially rule out the different scenarios for the origin of this mode. Nonetheless, the emergence of superconductivity in Ta doped TiSe$_2$, and its absence with similar levels of Nb doping may, in addition to the differences in the electronic densities of states, also lie in the difference in the way these dopants modify the phonon modes of the materials.

Finally, the electronic phase diagram as a function of temperature and doping level for many electron-doped 1T-TiSe$_2$ systems is summarized in **Figure 4**. For comparison to the present results for 1T-Ti$_{1-x}$Ta$_x$Se$_2$ and 1T-Ti$_{1-x}$Nb$_x$Se$_2$ the electronic phase diagrams for $Cu_x TiSe_2$ and $Pd_x TiSe_2$ are included in the figure. The CDW signature in the resistivity gets weaker with higher $x$ content in $Ti_{1-x}Ta_xSe_2$, and the CDW transition is driven down only slightly in temperature. This is different from the case in $Cu_x TiSe_2$, in which the CDW transition in $TiSe_2$ is driven down substantially in temperature with increasing Cu content, followed by the emergence of a superconducting state.[13] In the $Ti_{1-x}Ta_xSe_2$ system, the $x$ dependence of $T_c$ displays a dome-like shape that is broad in composition. The superconducting state appears for $x > 0.02$, going through a maximum $T_c$ of 2.2 K at $x = 0.2$, followed by a decrease of $T_c$ and then disappearance when at $x > 0.5$. Compared with $Cu_x TiSe_2$, the maximum $T_c$ in $Ti_{1-x}Ta_xSe_2$ is lower but the superconducting region is much broader. In addition, there is a significant boundary composition region ($0.02 < x < 0.2$) where superconductivity and CDW behavior may coexist. For the isoelectronic equivalent material $Ti_{1-x}Nb_xSe_2$, on the other hand, superconductivity does not appear for temperatures above 0.4 k for any of the materials. For the Pd-intercalated system, $Pd_x TiSe_2$, $T_c$ is low and is found for only a narrow composition range.[24]

**Conclusion**

We have found that $TiSe_2$ becomes superconducting when doped with Ta, a dopant which, consistent with a simple chemical picture, donates electrons to the conduction band. ARPES characterization of the resulting material shows that the Fermi surface is very similar to that seen for Cu-intercalated $TiSe_2$. The $T_c$ for the Ta doped case is a factor of 2 lower than that observed for Cu intercalation and is seen over a much wider range of electron doping concentrations. For chemically equivalent and chemically isoelectronic Nb doping, on the other hand, the phonon spectrum and the electronic system do appear to be significantly different. The smaller observed $\gamma$ is consistent with the observation that the Fermi surface and conduction band filling are significantly smaller in the Nb doped case than it is seen in the Ta doped case. That in itself would not obviously lead to the absence of superconductivity, since it emerges in other doped 1T-$TiSe_2$ systems at very low electron doping levels (i.e. $x \sim 0.02$), where the filling of the conduction band and thus the size of the electron Fermi surface is very small. The data overall imply that although chemically equivalent, the Nb dopant is not as effective in donating electrons into the conduction band of 1T-$TiSe_2$ as the Ta dopant is, even though it does weaken the CDW. Our comparison of the electronic phase diagrams for the different types of electron doping of 1T-$TiSe_2$ finds them to be quite different, clearly showing that how one chemically dopes electrons into the 1T-TiSe$_2$ system strongly matters. Although differences in the underlying electronic and phonon systems are observed, the fundamental reasons behind why Ta and Nb doping should lead to such differences remain obscure. The big difference between Nb and Ta doping in inducing superconductivity in the present material may have broad implications for doping-induced superconductivity in conventional electronic systems in general because failed attempts to introduce superconductivity in a material through chemical substitution may succeed if a different dopant is employed, or may be specific to the case of 1T-TiSe$_2$, which has certainly proven to be an unusual electronic material, and would be of interest for further study. We conclude by pointing out that while intercalation-induced superconductors such as 1T Cu$_x$TiSe$_2$ or Pd$_x$TiSe$_2$ may not be suitable for exfoliation and the fabrication of experimental devices due to the difficulty in cleaving TMDCs with intercalants that strongly bond the layers together, Ta-doped 1T-TiSe$_2$ is likely to be highly suitable for that purpose since the van der Waals bonding between MX$_2$ layers remains undisturbed in the superconducting material and exfoliation is expected to be relatively easy.

**Acknowledgements**

The materials synthesis and physical property characterization of this superconductor were supported by the Department of Energy, division of basic energy sciences, Grant DE-FG02-98ER45706. The single crystal structure determinations were supported by the Gordon and Betty Moore Foundation, EPiQS initiative, grant GBMF4412. The electron diffraction study at Brookhaven National Laboratory was supported by the DOE BES, by the Materials Sciences and Engineering Division under Contract DE-AC02-98CH10886, and through the use of the Center for Functional Nanomaterials. The ARPES experiments were performed under the LBNL and BNL grants DE-AC02-05CH11231 and DE-SC0012704. The Raman spectroscopy, AFM and mechanical exfoliation was supported by the National Science Foundation, Grant DMR-1410846.

**Author Contributions**

R.J.C and H.X.L conceived and designed the experiments and H.X.L performed the synthetic experiments and H.X.L and R.J.C analyzed and interpreted the chemical and transport data. W.W.X performed and analyzed the single diffraction data. H.X.L, E M.S and J.K performed and analyzed the XRD refinement data. J.T. and Y.M.Z performed and analyzed the electron diffraction data. I.P. and T.V. performed the ARPES measurements and analysis. G.O. and K.B. performed the Raman spectroscopy measurements and analyzed the Raman data. G.S.S. performed the XPS spectra measurements and analyzed the XPS data. H.X.L and R.J.C wrote the paper with input from all authors. All authors approved the content of the manuscript.

**Competing financial interests:** The authors declare no competing financial interest.

**Figures legends**

**Figure 1 Structural and chemical characterization of $Ti_{1-x}Ta_xSe_2$, $Ti_{1-x}Nb_xSe_2$** (a) Powder XRD patterns (Cu Kα) for selected samples ($TiSe_2$, $Ti_{0.8}Ta_{0.2}Se_2$ and $Ti_{0.8}Nb_{0.2}Se_2$) in this study. (b) Composition dependence of the room temperature lattice parameters for $Ti_{1-x}Ta_xSe_2$ (0 ≤ x ≤ 0.9) and $Ti_{1-x}Nb_xSe_2$ (0 ≤ x ≤ 0.7), compared with that of $Cu_xTiSe_2$ (0≤ x ≤ 0.11). Lattice parameters for $Cu_xTiSe_2$ were extracted from Ref 13. (e,f) XPS spectra of the Nb *3d* and Ta *4f* regions of $Ti_{0.8}Nb_{0.2}Se_2$ and $Ti_{0.8}Ta_{0.2}Se_2$. For comparison, the Nb *3d* and Ta *4f* spectra for undoped $2H-NbSe_2$ and $2H-TaSe_2$ are included in (c,d). The shifts in binding energy Δ compared to the absolute binding energy (i.e. Δ/B.E.) are very similar for both Nb and Ta dopants, showing them to be chemically equivalent when substituted in $1T-TiSe_2$. Electron diffraction in the [001] zones (g) and (i) $Ti_{0.8}Nb_{0.2}Se_2$ at room temperature (RT) 300 K and 89 K respectively. (h) and (j) the same two temperatures for $Ti_{0.8}Ta_{0.2}Se_2$. The CDW is present, visible due to its weak diffraction spots, in the Ta doped material at 89 K, but not in the Nb-doped material.

**Figure 2 Transport and specific heat characterization of the normal states and superconductuctivity.** (a) The temperature dependence of the resistivity ratio ($\rho/\rho_{300K}$) for polycrystalline $Ti_{1-x}Ta_xSe_2$ (0.02≤ x≤0.3). Inset: $d\rho/dT$ for $Ti_{1-x}Ta_xSe_2$ (0.05 ≤ x ≤ 0.25) in the low temperature region (1 - 3 K), showing the superconducting transition. (b) The temperature dependence of the resistivity ratio ($\rho/\rho_{300K}$) for polycrystalline $Ti_{1-x}Nb_xSe_2$ (0.02 ≤ x ≤ 0.7) Inset: enlarged view of the low temperature region (0.4 -3 K), showing the lack of a superconducting transition. (c) Temperature dependence of the resistivity of $Ti_{1-x}Nb_xSe_2$ as log ρ vs. log T. Red line is a fit to the 2D variable range hopping model at high temperatures. (d) Magnetic susceptibilities of $Ti_{0.8}Ta_{0.2}Se_2$ and $Ti_{0.8}Nb_{0.2}Se_2$ with applied field 5T. Inset: Hall measurement for $Ti_{0.8}Ta_{0.2}Se_2$ and $Ti_{0.8}Nb_{0.2}Se_2$. (e) Temperature dependence of the specific heat $C_p$ of $Ti_{0.8}Ta_{0.2}Se_2$ measured under magnetic fields of 0 T and 5 T, presented in the form of $C_p/T$ vs $T^2$ (main panel) and $C_{el}/T$ vs T (inset). The green line shows the equal area construction to determine $\Delta C/\gamma T_c$. The red line shows the fit of the specific heat in the range 2 - 10 K at 5 T. (f) Temperature dependence of the specific heat $C_p$ of $Ti_{0.8}Nb_{0.2}Se_2$ measured under a magnetic field of 0 T, presented in the form of $C_p/T$ vs $T^2$.

**Figure 3. Probing the electronic structure and phonon spectra of the doped $1T-TiSe_2$ materials.** Performed on the (001) crystal surface **ARPES measurements at 15 K** and Raman spectra at 300 K. ARPES-determined Fermi surface cuts at the border of the Brillouin zone in the plane containing the high-symmetry points A, L, and H at $k_c \sim \pi/c$ for (a) $Ti_{0.85}Ta_{0.15}Se_2$ and (b) $Ti_{0.85}Nb_{0.15}Se_2$, showing the qualitatively smaller Fermi surface for the case of Nb doping. (c) and (d) The ARPES-determined band dispersion across the electron pocket at L for (c) $Ti_{0.85}Ta_{0.15}Se_2$ and (d) $Ti_{0.85}Nb_{0.15}Se_2$ respectively, again showing the smaller filling of the electron pocket in the Nb-doping case. (e) and (f): The bands forming the hole pockets in the center of the $k_c$-projected Brillouin zone for (e) $Ti_{0.85}Ta_{0.15}Se_2$ and (f) $Ti_{0.85}Nb_{0.15}Se_2$ respectively. (g) and (h): The band dispersions along Γ-M at $k_c \sim 0$ for pristine $1T-TiSe_2$ and Nb-doped $TiSe_2$, respectively. Spectra were taken at 15 K using photon excitation of 78 eV (a)-(f) and 95 eV (g)-(h). (i) Raman spectra reveal no modification of the phonons of $1T-TiSe_2$ by Nb doping and that Ta doping shifts the higher energy $A_{1g}$ mode to lower energies

and induces a new mode at 213 cm$^{-1}$. (j) Fit of the Raman spectrum in A$_{1g}$ region of Ti$_{0.85}$Ta$_{0.15}$Se$_2$ clearly showing the existence of the new mode and the blue shift of the original A$_{1g}$ phonon.

**Figure 4 The electronic phase diagram of the superconducting 1T-TiSe$_2$ system.** The electronic phase diagrams for Cu$_x$TiSe$_2$, Pd$_x$TiSe$_2$, Ti$_{1-x}$Ta$_x$Se$_2$ and Ti$_{1-x}$Nb$_x$Se$_2$ are shown as a function of Cu, Pd, Ta or Nb content *x*. All the nominally electron-doped systems are different. Superconductor parameters for Cu$_x$TiSe$_2$ and Pd$_x$TiSe$_2$ were extracted from Refs. 13 and 24, respectively.

**Table S1.** Single crystal crystallographic data for Ti$_{0.81}$Nb$_{0.19(1)}$Se$_2$ and Ti$_{0.88}$Ta$_{0.12(1)}$Se$_2$ at 100(2) K.

**Table S2.** Atomic coordinates and equivalent isotropic displacement parameters of Ti$_{0.81}$Nb$_{0.19}$Se$_2$ and Ti$_{0.88}$Ta$_{0.12}$Se$_2$ at 100 K. U$_{eq}$ is defined as one-third of the trace of the orthogonalized U$_{ij}$ tensor (Å$^2$).

**Figure 1S.** XPS spectra of the Se *3d* regions of Ti$_{0.8}$Nb$_{0.2}$Se$_2$ and Ti$_{0.8}$Ta$_{0.2}$Se$_2$. For comparison, the Se *3d* spectrum for undoped 2H-NbSe$_2$ and 2H-TaSe$_2$ are included in (g,h)

**Figure 2S.** XPS spectra of the Ti *2p* regions of Ti$_{0.8}$Nb$_{0.2}$Se$_2$ and Ti$_{0.8}$Ta$_{0.2}$Se$_2$.

**Figure 3S. The upper critical field characterization of Ti$_{1-x}$Ta$_x$Se$_2$.** Low temperature resistivity at various applied fields for (a) Ti$_{0.8}$Ta$_{0.2}$Se$_2$ and (b) Ti$_{0.85}$Ta$_{0.15}$Se$_2$. Inset shows the temperature dependence of the upper critical field (H$_{c2}$).

Figure 1

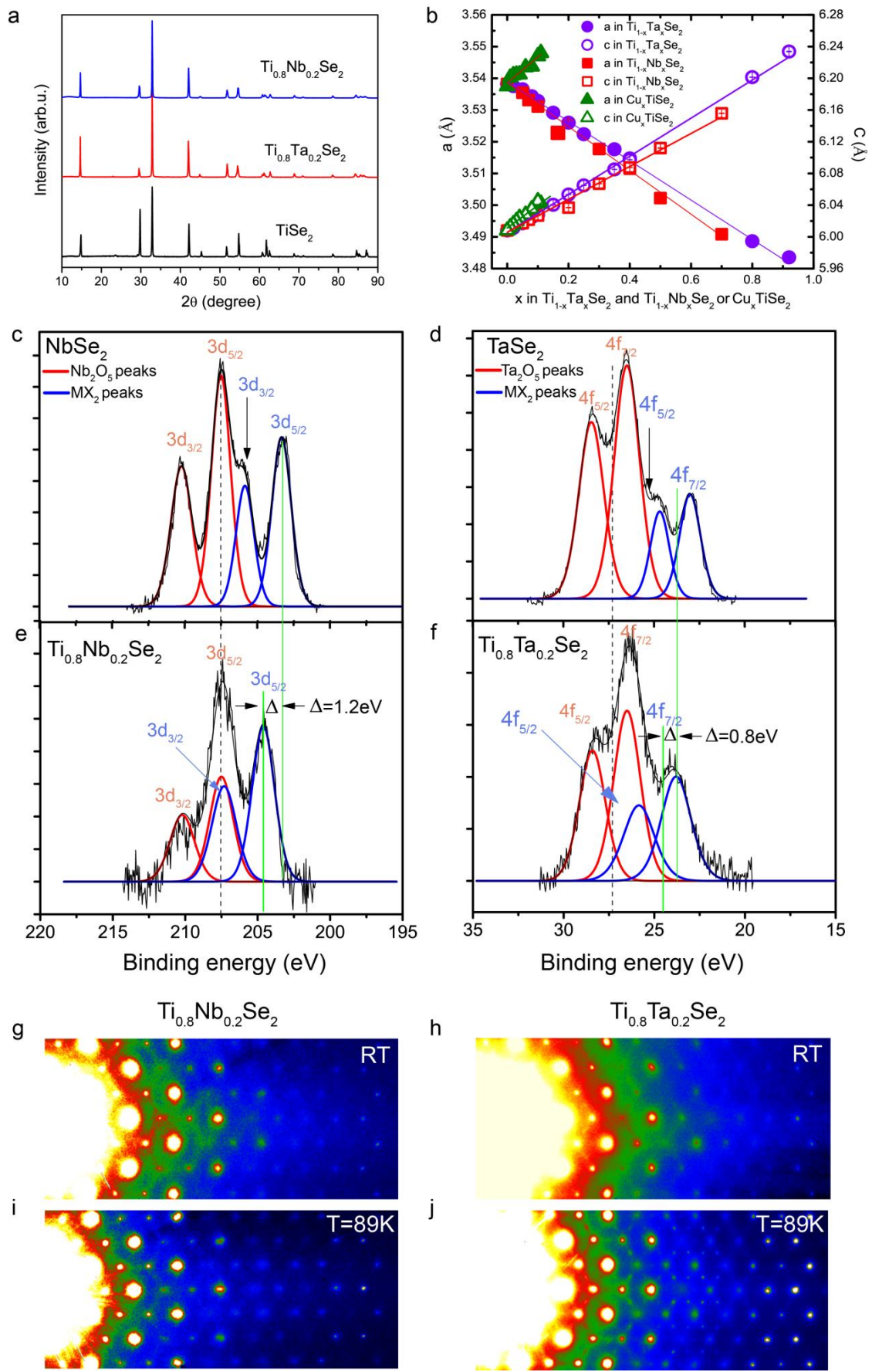

Figure 2

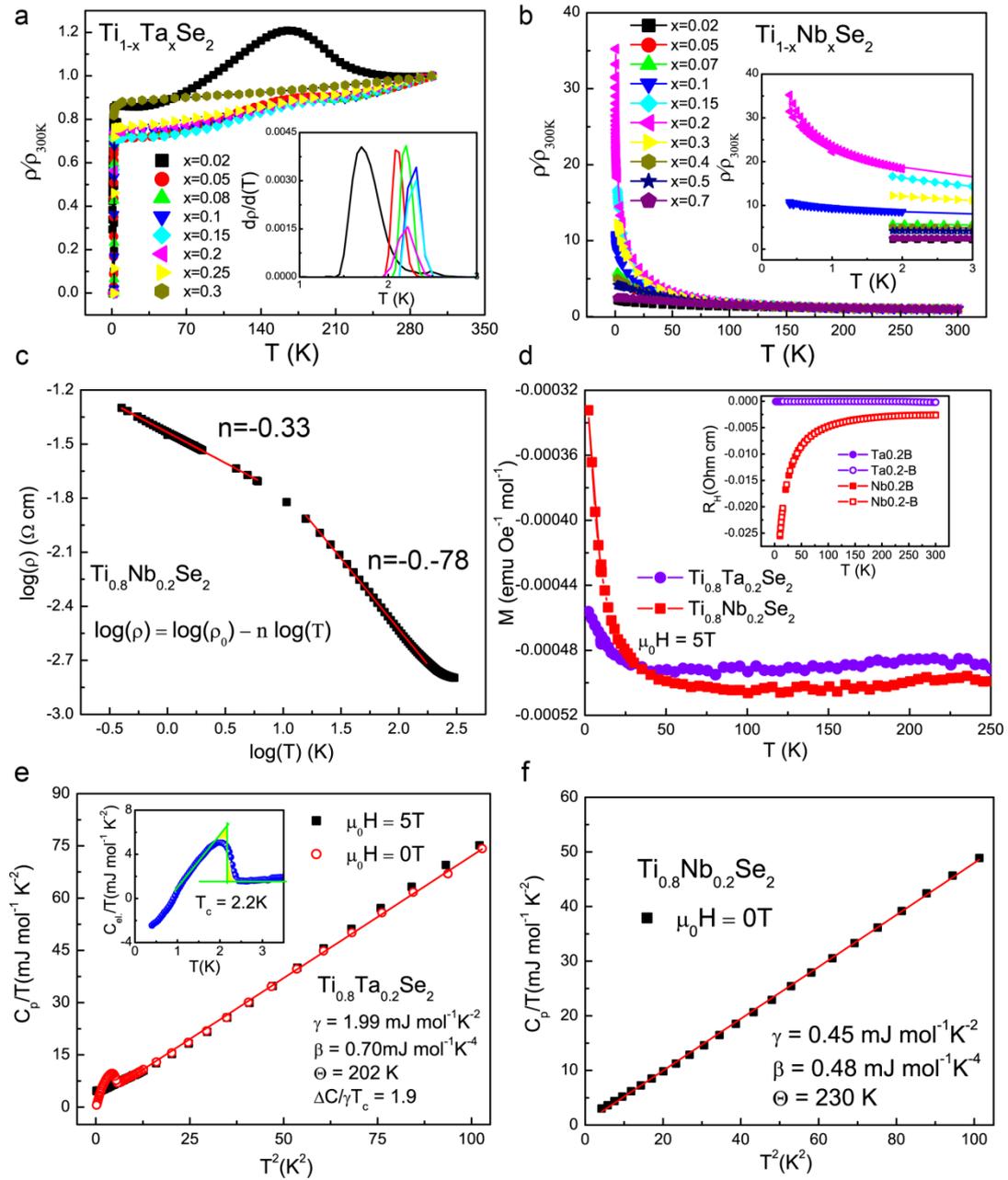

**Figure 3**

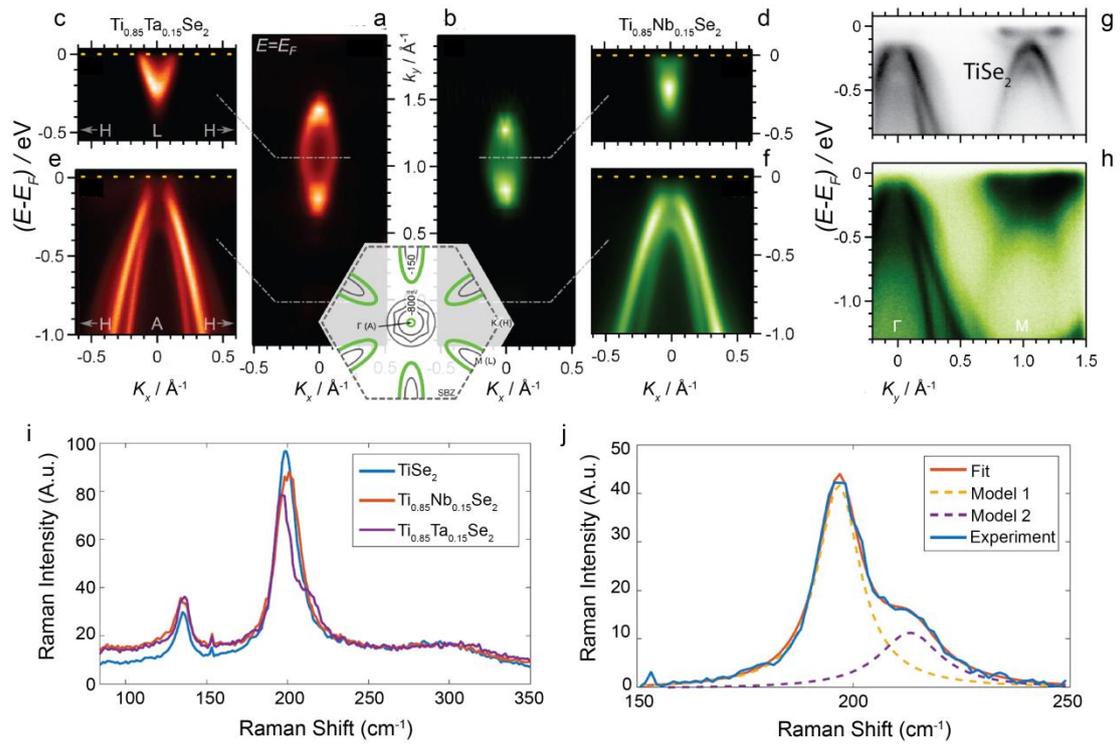

**Figure 4**

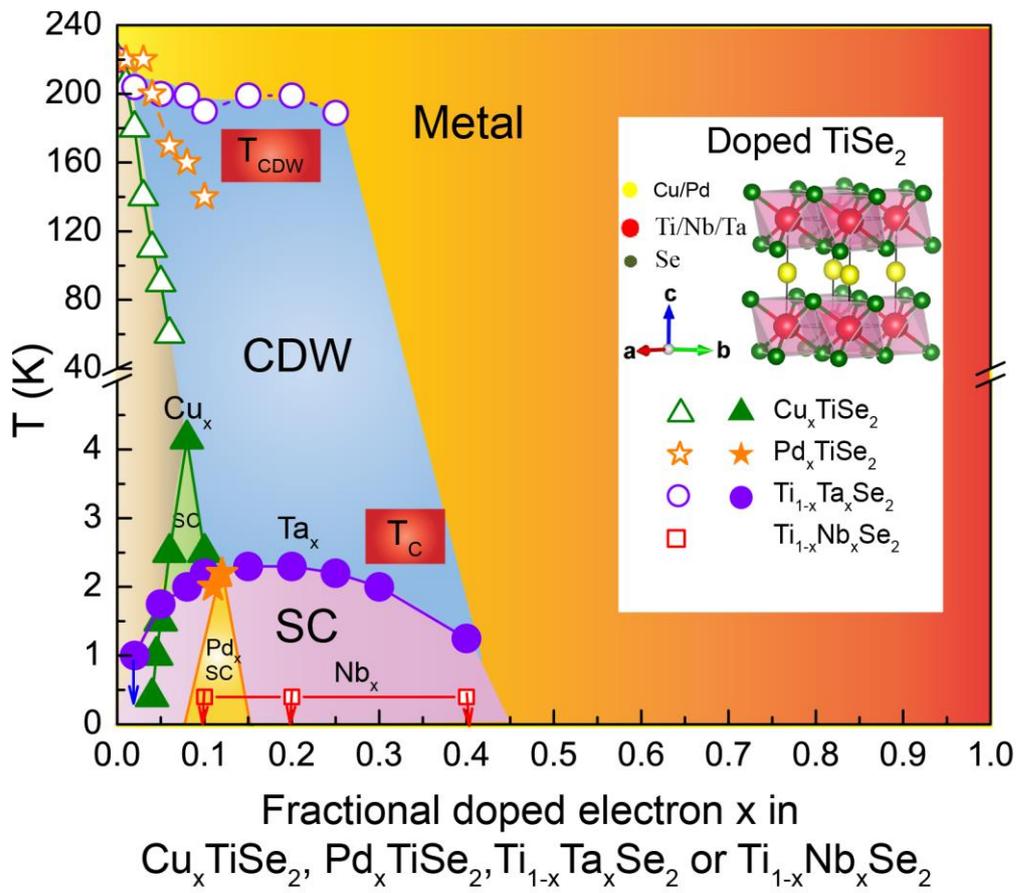

**Table of Content**

**Differences in chemical doping matter -**

**Superconductivity in $Ti_{1-x}Ta_xSe_2$ but not in $Ti_{1-x}Nb_xSe_2$**

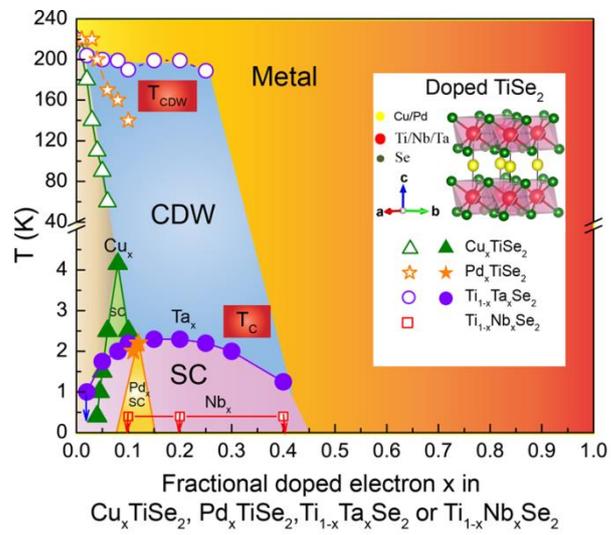

Supplementary information

# Differences in chemical doping matter - Superconductivity in $Ti_{1-x}Ta_xSe_2$ but not in $Ti_{1-x}Nb_xSe_2$

Huixia Luo[1,*], Weiwei Xie[1], Jing Tao[2], Ivo Pletikosic[2,3], Tonica Valla[2], Girija S. Sahasrabudhe[1], G. Osterhoudt[4], Erin Sutton[4], K. Burch[4], Elizabeth M. Seibel[1], Jason W. Krizan[1], Yimei Zhu[2] and R. J. Cava[1]

[1]*Department of Chemistry, Princeton University, Princeton, NJ 08544, USA*

[2]*Condensed Matter Physics and Materials Science Department, Brookhaven National Lab, Upton, New York 11973, USA*

[3]*Department of Physics, Princeton University, Princeton, NJ 08544, USA*

[4]*Department of Physics, Boston College, Boston MA*

*huixial@princeton.edu

**Crystal Structure analyses**

X-ray diffraction intensity data were collected at 100 K on a Bruker Apex Photon diffractometer with Mo radiation $K\alpha_1$ ($\lambda$ = 0.71073 Å) or Cu radiation $K\alpha_1$ ($\lambda$ = 1.54098 Å). Data were collected over a full sphere of reciprocal space with 0.5° scans in ω with an exposure time of 10s per frame. The 2θ range extended from 4° to 60°. The SMART software was used for data acquisition. Intensities were extracted and corrected for Lorentz and polarization effects with the SAINT program. Empirical absorption corrections were accomplished with SADABS, based on modeling a transmission surface by spherical harmonics employing equivalent reflections with I > 2σ(I).[S1-4] Within the SHELXTL package, the crystal structures were solved using direct methods and refined by full-matrix least-squares on $F^2$.[S3]

**References**
S1. Sheldrick, G. M. (2001) .SADABS, University of Gottingen, Gottingen,Germany.
S2. Sheldrick, G. M. (2008) A short history of SHELX. Acta Crystallogr. A 64:112.
S3. SHELXTL, version 6.10, Bruker AXS Inc.: Madison, WI, 2000.
S4. Momma K, Izumi F (2011) VESTA 3 for three-dimensional visualization of crystal, volumetric and morphology data. J. Appl. Crystallogr. 44:1272.

**Table S1.** Single crystal crystallographic data for Nb and Ta doped TiSe$_2$ at 100(2) K.

| Refined Formula | Ti$_{0.81(1)}$Nb$_{0.19}$Se$_2$ | Ti$_{0.879(4)}$Ta$_{0.121}$Se$_2$ |
|---|---|---|
| F.W. (g/mol); | 214.37 | 221.79 |
| Space group; Z | P3-m1(No.164); 1 | P3-m1(No.164); 1 |
| a (Å) | 3.5217(1) | 3.5180(2) |
| c (Å) | 6.0443(3) | 6.0093(4) |
| V (Å$^3$) | 64.920(5) | 64.409(8) |
| Absorption Correction | Numerical | Numerical |
| Radiation | Cu | Mo |
| Extinction Coefficient | None | None |
| θ range (deg) | 7.327-61.839 | 3.390-30.045 |
| No. reflections; $R_{int}$ | 617;0.0280 | 810;0.0138 |
| No. independent reflections | 53 | 96 |
| No. parameters | 8 | 8 |
| $R_1$; $wR_2$ (all $I$) | 0.0341; 0.0438 | 0.0175; 0.0404 |
| Goodness of fit | 1.294 | 1.392 |
| Diffraction peak and hole (e$^-$/Å$^3$) | 0.636;–1.109 | 0.953; –1.052 |

**Table S2.** Atomic coordinates and equivalent isotropic displacement parameters of Nb and Ta doped TiSe$_2$ at 100(2) K. U$_{eq}$ is defined as one-third of the trace of the orthogonalized U$_{ij}$ tensor (Å$^2$).

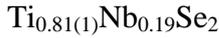

Ti$_{0.81(1)}$Nb$_{0.19}$Se$_2$

| Atom | Wyckoff. | Occupancy. | x | y | z | U$_{eq}$ |
|---|---|---|---|---|---|---|
| Ta/Ti1 | 1a | 0.81(1)/0.19 | 0 | 0 | 0 | 0.0124(13) |
| Se2 | 2d | 1 | 1/3 | 2/3 | 0.2581(2) | 0.0104(7) |

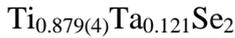

Ti$_{0.879(4)}$Ta$_{0.121}$Se$_2$

| Atom | Wyckoff. | Occupancy. | x | y | z | U$_{eq}$ |
|---|---|---|---|---|---|---|
| Ta/Ti1 | 1a | 0.879(4)/0.121 | 0 | 0 | 0 | 0.0059(5) |
| Se2 | 2d | 1 | 1/3 | 2/3 | 0.2577(1) | 0.0043(2) |

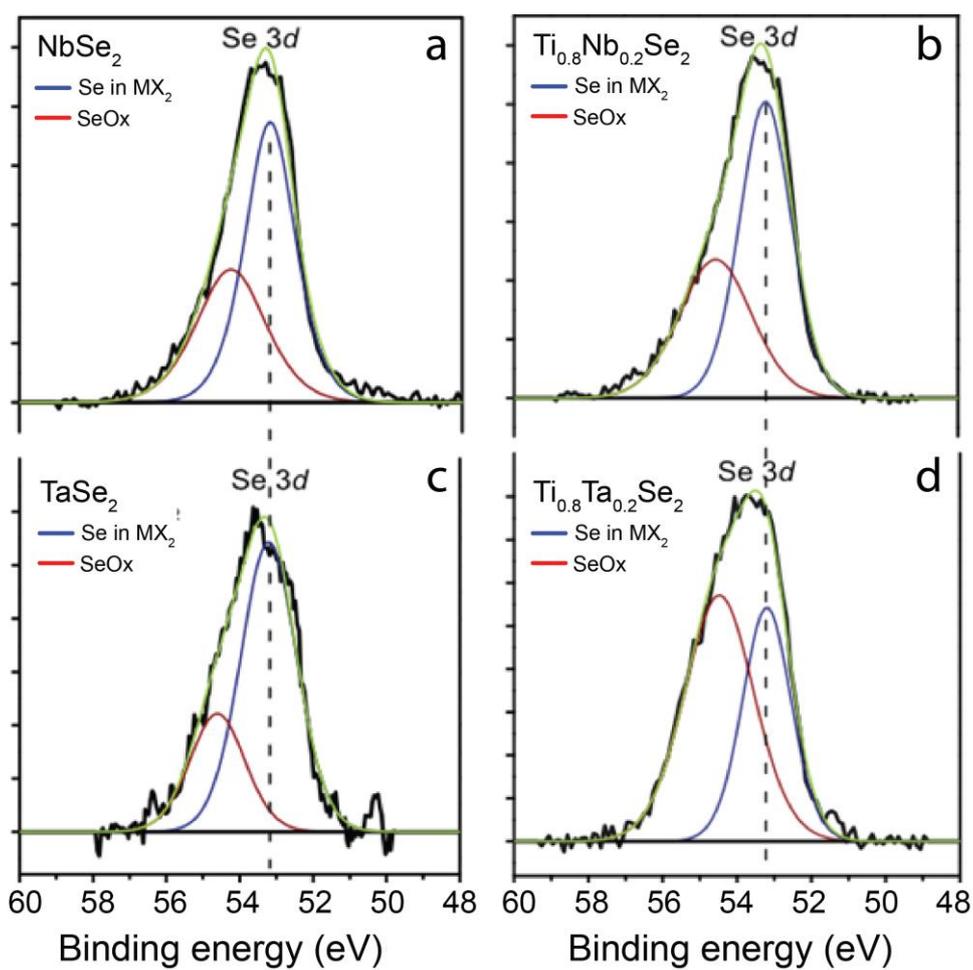

**Figure 1S**. XPS spectra of the Se $3d$ regions of $Ti_{0.8}Nb_{0.2}Se_2$ and $Ti_{0.8}Ta_{0.2}Se_2$ (c,d). For comparison, the Se $3d$ spectra for undoped 2H-NbSe$_2$ and 2H-TaSe$_2$ are included in (a,b)

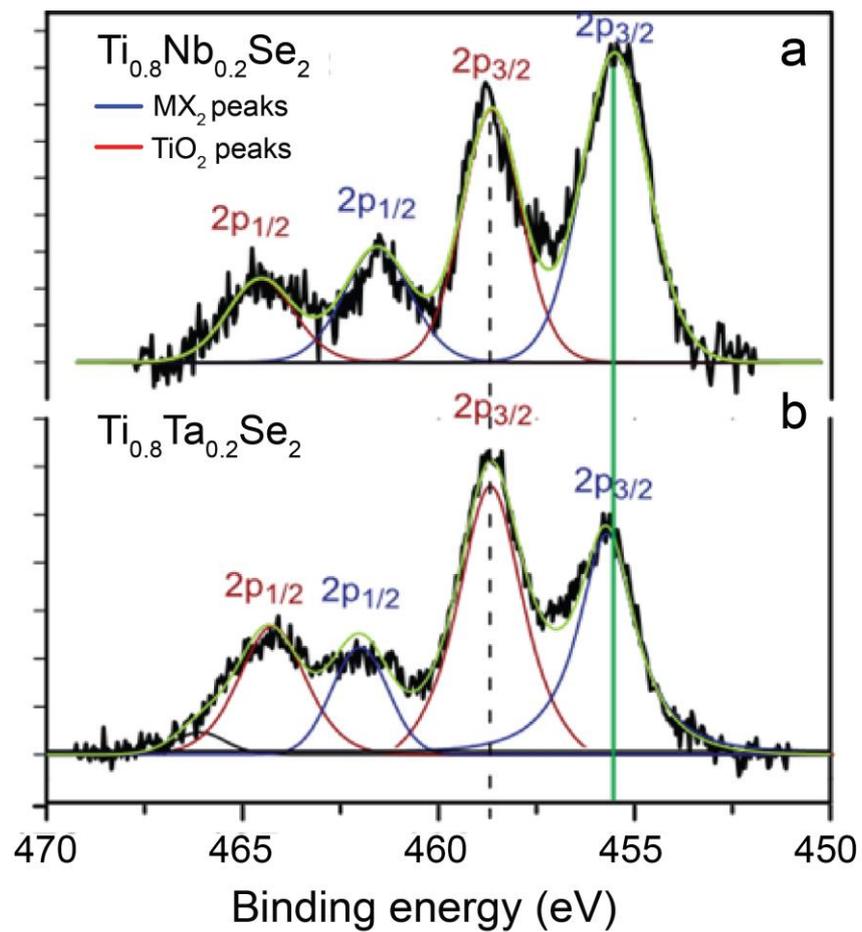

**Figure 2S**. XPS spectra of the Ti *2p* regions of $Ti_{0.8}Nb_{0.2}Se_2$ and $Ti_{0.8}Ta_{0.2}Se_2$.

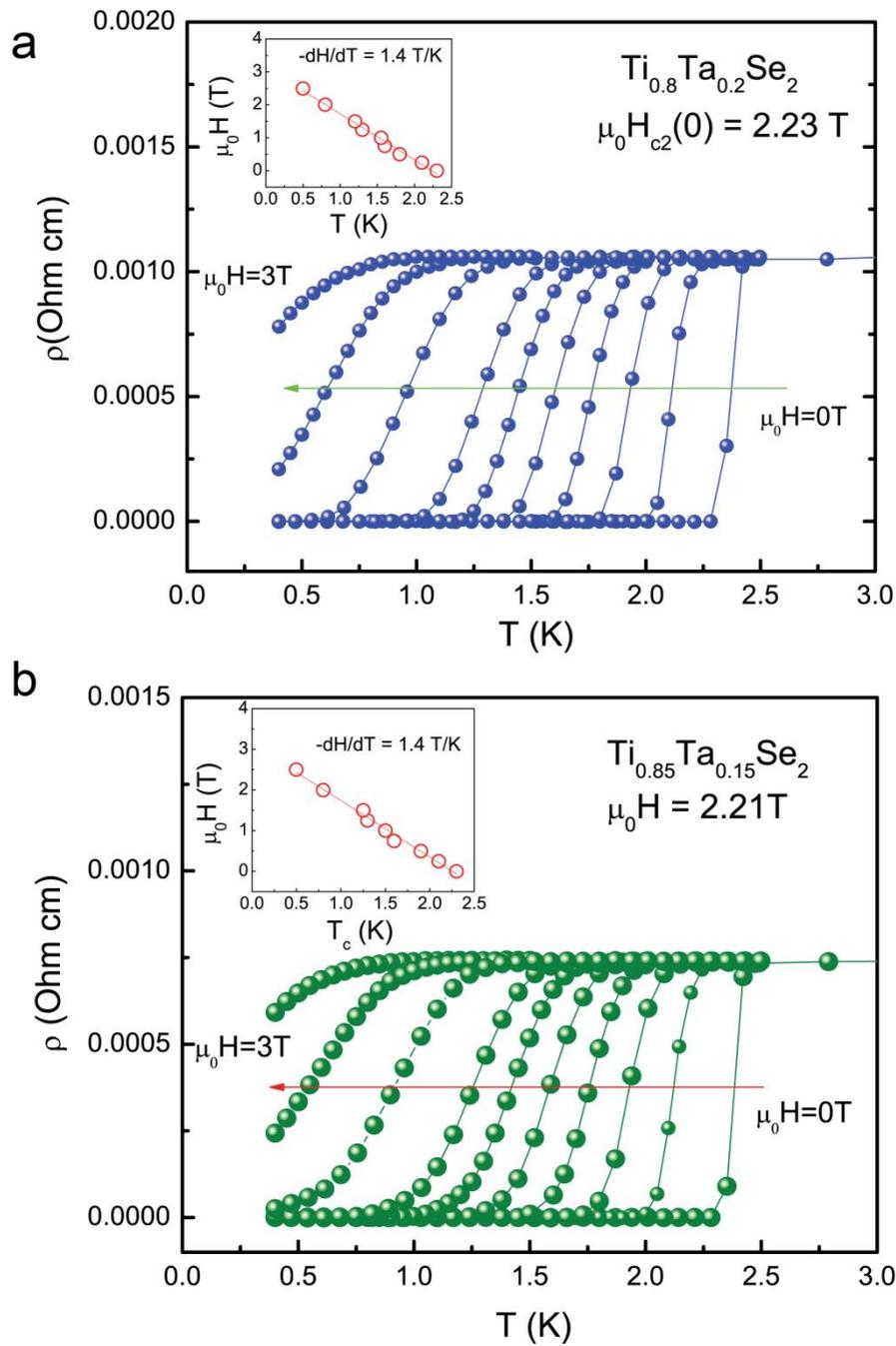

**Figure 3S. The upper critical field characterization of $Ti_{1-x}Ta_xSe_2$.** Low temperature resistivity at various applied fields for (a) $Ti_{0.8}Ta_{0.2}Se_2$ and (b) $Ti_{0.85}Ta_{0.15}Se_2$. Inset shows the temperature dependence of the upper critical field ($H_{c2}$).